\documentclass[10pt, onecolumn]{IEEEtran}

\usepackage{fancyhdr}
\fancypagestyle{empty}{
\fancyhf{}

\cfoot{This work has been submitted to IEEE Control Systems Letters for review.}}
\usepackage{amsmath,amscd,amssymb,amsgen,amsfonts,amsbsy}
\usepackage{mathrsfs}
\usepackage[utf8x]{inputenc}
\usepackage{color}
\usepackage{textcomp}
\usepackage{float}
\usepackage{latexsym,graphicx}

\usepackage{tikz}
\usetikzlibrary{automata,arrows,positioning,calc}
\usepackage{url}
\usepackage{cite}
\usepackage[ruled,lined]{algorithm2e}
\SetAlFnt{\small}
\usepackage{breqn}

\newcommand{\remove}[1]{}

\newcommand{\comments}[1]{}

\newtheorem{lemma}{Lemma}

\newtheorem{theorem}{Theorem}
\newtheorem{remark}{Remark}

\newtheorem{definition}{Definition}
\makeatother

\title{Restless Bandits with Constrained Arms: Applications in Social and Information Networks\footnote{This work has been submitted to IEEE Control Systems Letters for review.}}
\author{
 \begin{tabular}{ccc}
 Varun Mehta, Rahul Meshram, Kesav Kaza and S.~N.~Merchant \\ 
      Department of Electrical Engineering          \\
    IIT Bombay, Mumbai INDIA. \\     
  \end{tabular}}

\begin{document}
%

\maketitle
\begin{abstract}
We study a problem of information gathering in a social network with
dynamically available sources and time varying quality of
information. We formulate this problem as a restless multi-armed bandit (RMAB). 
In this problem, information quality of a source corresponds
to the state of an arm in RMAB. The decision making agent does not know the quality of
information from sources a priori. But the agent maintains a belief
about the quality of information from each source. This is a problem of
RMAB with partially observable states.  The objective of the agent is to
gather relevant information efficiently from sources by contacting
them.  We formulate this as a infinite horizon discounted reward
problem, where reward depends on quality of information.  We study
Whittle's index policy which determines the sequence of play of arms
that maximizes long term cumulative reward. We illustrate  
the performance of index policy, myopic policy and
 compare with uniform random policy through numerical simulation.
\end{abstract}


\IEEEpeerreviewmaketitle

\section{Introduction}
%

Suppose there is an agent in a social or information network. The
agent has connections to $N$ neighbors which are its information
sources.  The agent needs information for its use and it gathers this
information through its sources at regular intervals.  In each
interval, the agent can contact only $M<N$ of its sources for
information. The information provided by a source may be either
relevant ($1$) or non-relevant ($0$) to the agent.  So, there are two
states, say $\{0,1\},$ which corresponds to the information quality.
However, the agent does not know a priori whether a certain source has
relevant information. Relevance of the information received becomes
apparent to the agent at end of the interval after processing
it. The assumption that information quality is binary comes
  from the following consideration. Agents who want to form informed
  opinion on a given matter will first decide to access relevant
  material from a source which they believe to be accurate. Also, in
  some scenarios relevant and irrelevant states may be interpreted as
  truth or falsehood. The agent however knows that the information
quality of a source varies in a Markovian manner and also knows its
Markov matrix. Assume that the sources are self interested
  entities. Their current relevance/truthfulness depends on their
  previous state. The reward from relevant information is high and
non-relevant information is low.

 Further, in a given interval each source may or may not be
 available. However, an unavailable source may be leveraged through an
 additional cost. Hence the immediate reward that such information
 gives may be lower. This is a situation where the choice of sources
 might effect their future availability and information quality of
 the sources.  The agent here needs a policy to choose which sources
 it must contact in each interval along the time line so that its
 cumulative reward is maximized.  

A dynamic information sourcing problem such as above is sequential
decision problem where a current decision impacts future rewards.
Such sequential decision problems can be modeled using restless
multi-armed bandits (RMAB) (see \cite{Gittins11,Whittle88}).
An RMAB is an agent with $N$ arms and each arm can be in one of finite
states. The states of arms evolve along Markov chains whose transition
probabilities are known to the agent. The agent can pull $M<N$ arms at
a time. Reward of pulling an arm depends on its state. The agent want
to find a policy to make these arm choices in each slot to maximize
its long term cumulative reward.

We now briefly review some related work.  Information gathering in
social networks has been a topic of interest in context of business
and management decisions. \cite{Borgatti03,Goyal17} study the impact
of parameters such as perception, cost, timely access, etc., on choice
of sources for information seeking in a large organization.  An
empirical study on journalists' use of social media for sourcing
information has been conducted by \cite{Lariscy09}. Further, automated
monitoring of social media to build geo-spatial awareness during
disasters is attempted in \cite{Yin12}.  In general, in information intensive
applications such as governance, public relations, journalism etc.,
both manual and automated information gathering happen in a
sequential manner. Sequential decision making models in information
networks have been studied under different scenarios in 
\cite{Papanastasiou17,Drakopoulos13}. 



In seminal paper \cite{Whittle88}, author introduced RMAB problem and
proposed a heuristic index based policy; such index policy is now
referred to as Whittle index policy. RMAB assumes that the model of
system state variations is known.  The Whittle index based policies
also studied for opportunistic communication systems in
\cite{LiuZhao10,Meshram16}, where authors studied partially observable
model. The Whittle index policies are popular due to they are
asymptotically optimal. In \cite{LiuZhao10}, index policy is shown to
be optimal for certain model parameter.  In \cite{Meshram16}, a hidden
Markov RMAB is studied which generalizes the work of
\cite{LiuZhao10}. The myopic policy was also investigated for restless
bandits in \cite{Ahmad09} and it is shown to be optimal under certain
model assumptions.
%

All the above models of RMAB assume that each arm is always available
to play, and the decision in each slot is whether to play or not play
an arm. However, availability of arms can be dynamic. Multi-armed
bandit problems with dynamic availability constraints have been
studied for machine-repair problem in \cite{Dayanik02}, where, if the
machine breaks down, then it will be available in next time slot with
some probability after getting repaired. This model assumes that the
state is observable and the authors analyze index-type policy for
rested bandits.

In this paper, we formulate the problem of an agent gathering
information from neighbors in a social network as RMAB with
constrained arms (each arm being a dynamically available source).  We
further study Whittle index policy and myopic policy. We show that,
for a single-armed bandit with constrained arm, the optimal policy is
of a threshold-type, and the arm is indexable. We next devise the
algorithm to compute Whittle index for each arm and this algorithm is
based on two timescales stochastic approximations scheme. We finally illustrate performance of this scheme via numerical example.

We next describe the system model.
\section{Preliminaries and Model Description}
\label{sec:model}
There is an agent in a social or information network. The agent has
connections to $N$ neighbors which are its information sources.  The
agent can contact only $(M<N)$  neighbors for information.  The
system is assumed to be time slotted and it is indexed by $t.$
The quality of information available at the source represented by a
Markov chain with state space $\{0,1\}.$ Let $X_n(t)$ denote the state
corresponding to information quality of source $n$ at beginning of
time slot $t,$ $X_n(t) \in \{0,1\}.$
We suppose that each source has dynamic availability i.e. in a given
slot it may or may not be available.  When a source is not available,
it may be leveraged to provide information by incurring an additional
cost.
 Let $Y_n(t) \in \{ 0,1\}$ represent the availability of the source
 $n$ in time slot $t.$
Since the agent contacts $M$ sources out of $N$ in each time slot to
gather information, we define $A_n(t) \in \{0,1 \}$ as the action in
slot $t,$ where $ A_n(t) = 1$ if source $n$ is contacted in slot $t,$ 
and $ A_n(t) = 0$ otherwise. 
%
%
We can have $A_n(t)=1$ in both available and unavailable scenarios.
The state of arm $n,$ i.e., $X_n(t)$ changes at the end of time slot
$t$ according to transition probabilities that depend on $A_n(t),$
$Y_n(t)$ and it is defined as follows.

{{
\begin{eqnarray*}
P_{ij}^n(y,a) := \Pr \{ {X_n}(t + 1) = j~|~{X_n}(t) = i,{Y_n}(t) = y,{A_n}(t) = a\}.
\end{eqnarray*}
}}
If source $n$ is contacted in slot $t,$ then quality of information
from source $n$ is known exactly at the end of slot, i.e., state of
source is known exactly. Also, the agent makes a binary observation
$Z_n^y(t)$ about source $n$ when that is contacted. Hence, we define
$Z_n^{y}(t) := 1$ if information from source $n$ is relevant, and
$Z_n^{y}(t) := 0$ otherwise.
%
Let $\rho_{n}(i,y)$ be the probability of $Z_n^{y}(t) =1$ given that
$X_n(t) = i,$ $Y_n(t) = y$ and $A_n(t) =1.$

{{
\begin{eqnarray*}
  \rho_n(i,y) := \Pr{\left(Z_n^{y}(t) = 1~|~X_n(t) = i, Y_n(t) = y, A_n(t) =1 \right) }.
 \end{eqnarray*}
}}
We assume that $ \rho_{n}(0,y) = 0$ and $\rho_{n}(1,y) = 1$ for all $y
\in \{0,1\}.$ When source $n$ is not used, the agent do not know the
quality of information, hence state of source $n$ is unobservable.
Hence, the agent maintains a belief $\pi_n(t)$ about
the state of source $n.$ Here, belief is the probability that the
source is in state $0$ given all past availability, actions,
observations and given as
\begin{eqnarray*}
\pi_n(t) = \Pr{\left( X_n(t) = 0~|~\left( Y_n(s) = y_s,A_{n}(s)
  ,Z_{n}^{y_s}(s) \right)_{s=1}^{t-1} \right)}.
\end{eqnarray*}
We now define the reward as measure of the quality of information from
different sources.  When the agent uses source $n,$ it obtains reward
from the information it receives. This reward depends on current state
of that source and availability of that source.
Let $R_n^{a}(i,y)$ be the reward obtained from using source $n$ given
that $X_n(t) = i,$ $Y_n(t) = y,$ $A_n(t) =a,$ and it is as follows. 
\begin{eqnarray*}
 & R_n^{1}(i,1)=r_{n,i},  \hspace{0.2in} R_n^{1}(i,0) = \eta_{n,i}, \\
 &\hspace{-0.14in} R_n^{0}(i,1) = 0, \hspace{0.35in}  R_n^{0}(i,0) = 0.
\end{eqnarray*}
We further assume that $r_{n,0} = \eta_{n,0} = 0,$ no reward from
source $n$ if it has $X_n(t) = 0.$ Also, we suppose $r_{n,1} >
\eta_{n,1},$ for all $n.$ This implies that an unavailable source may
be leveraged through an additional cost. Hence, the immediate reward is
lower than when source is available.
%
However, agent knows that the availability of sources is dynamic.
This dynamic availability of each source $n$ is modeled stochastically
as probability of availability $\theta_n^a = \Pr{ \left({Y_n}(t+1) =
  1|A_n(t) = a \right)}.$ Thus availability of a source varies
according to Bernoulli distribution with parameter $\theta_n^a .$
This is known to the agent.
%
%
Let $H_t$ denote the history up to time $t$,
\begin{eqnarray*}
H_t := \left(Y_n(s) = y_s,A_{n}(s) ,Z_{n}^{y_s}(s) \right)_{1 \leq n \leq N, 1 \leq s < t}.
\end{eqnarray*}
We can  describe the state of source $n$ at time $t$  by 
$S_n(t) = (\pi_n(t), Y_n(t)) \in [0,1] \times \{0,1\}.$ 
$(S_1(t), \cdots S_N(t))$ is the state information of all the sources at the 
beginning of time slot $t.$ 
The expected reward from using source $n$ at time $t$ given that
$Y_n(t) = y$ is
\begin{equation*}
\widetilde{R}_n^{1}(\pi_n(t),y) = \pi_n(t) R_n^{1}(0,y) + (1-\pi_n(t) R_n^{1}(1,y).
\end{equation*}

In each slot, agent uses exactly $M$ sources. Let $\phi(t)$ is the
policy of agent such that $\phi(t): H_t \rightarrow \{1, \cdots,N\}$
maps the history to $M$ sources at each slot $t.$ Let
\begin{eqnarray*}
A_n^{\phi}(t) = 
\begin{cases}
1 & \mbox{if $n \in \phi(t),$ } \\
0 & \mbox{if $n \notin \phi(t),$}
\end{cases}
\end{eqnarray*} 
and $\sum_{n=1}^{N} A_n^{\phi}(t) = M.$\\

We are now ready to define the
infinite horizon discounted reward under policy $\phi$ for initial
state information $(\underline{\pi}, \underline{y}),$
$\underline{\pi}=(\pi_1(1), \cdots, \pi_N(1))$ and $\underline{y} =
(y_1(1), \cdots, y_N(1)).$ It is given by
{{
\begin{eqnarray*}
V_{\phi}(\underline{\pi}, \underline{y}) = 
\mathrm{E}^{\phi}\left({\sum_{t=1}^{\infty} \beta^{t-1} 
\left[ \sum_{n=1}^{N} 
A_n^{\phi}(t) \widetilde{R}_n^{1}(\pi_n(t), Y_n(t) )
\right]
}\right).
\end{eqnarray*}  }}
Here, $\beta$ is discount parameter, $0<\beta < 1.$ 
Then {{ 
\begin{eqnarray}
& \phi^{*} = \arg \max_{\phi} V_{\phi}(\underline{\pi}, \underline{y}) \nonumber \\ 
& s.t. \sum_{n=1}^{N} A_n^{\phi}(t) = M, \underline{\pi} \in [0,1]^N, \underline{y} \in \{0,1\}^N.
\label{eqn:opt1}
\end{eqnarray} }}
The optimization problem \eqref{eqn:opt1} is a restless multi-arm
bandit problem with availability constraints. Here, each source will
correspond to an arm.  The state of information quality of source $n$
and its availability represent the state $S_n(t)=(\pi_n(t),Y_n(t))$ of
an arm $n$. This is a generalized version of restless multi-arm
bandits with partially observable states and availability
constraints. This problem is known to be
PSPACE-hard,\cite{Papadimitriou99}. In this paper
we consider index based policies.  In such index polices, the
dimensionality of the problem is reduced by calculating the index for
each arm separately. The $M$ arms with highest indices are played at
each time slot. That is, the agent uses $M$ sources with highest
indices.
To use index policies, one requires to study relaxed version of
optimization problem~\eqref{eqn:opt1}, where a subsidy $w$ is
introduced for not playing arm (not using source by agent), see
\cite{Whittle88,Gittins11}.  We first analyze agent with a
single-armed bandit (a single source scenario) in next section.
\section{A Single-armed bandit problem}
\label{sec:sab}
For notation convenience, we will drop the subscript $n.$ We use the
terms arm and source interchangeably. In view of subsidy $w,$ we can
rewrite optimization problem~\eqref{eqn:opt1} for a single-armed
bandit as follows.

{{
\begin{dmath*}
\phi^{*} = \arg\max_{\phi} V_{\phi}(\pi, y) \\
{V_{\phi}(\pi, y) = 
\mathrm{E}^{\phi}\left(\sum_{t=1}^{\infty} \beta^{t-1} 
\left[  
A^{\phi}(t) \widetilde{R}^{1}(\pi(t), Y(t) ) +w(1-A^{\phi}(t))\right]
\right) }
\label{eqn:opt2}
\end{dmath*}
} } 
for initial belief $\pi \in [0,1]$ and availability $y \in \{0,1\}.$
Here, action $A(t)$ under policy $\phi$ is
{{
\begin{eqnarray*}
A^{\phi}(t) = 
\begin{cases}
1 & \mbox{if $\phi(t) = 1,$ } \\
0 & \mbox{if $\phi(t) =0.$}
\end{cases}
\end{eqnarray*} 
}}
We further simplify the model and assume that $P_{00}(y,a)= \mu _0$ and
$P_{10}(y,a)= \mu _1$ for $a,y \in \{0,1\}.$\footnote{In general,
  Markov model for source availability and unavailability could be
  different.} 
%
%
Recall that $\pi (t)= \Pr (X(t)=0 | H_t)$ and using the Bayes rule, we
update the belief $\pi (t+1)$ in following manner.

{{
\begin{eqnarray*}
\pi(t+1) = 
\begin{cases}
\mu_1 & \mbox{if $A(t) =1,$ $Y(t) = y,$ and $Z^{y}(t) = 1,$ } \\
\mu_0 & \mbox{if $A(t) =1,$ $Y(t) = y,$ and $Z^{y}(t) = 0,$ } \\
\Gamma(\pi(t)) & \mbox{if $A(t) = 0,$ and $Y(t) = y,$}
\end{cases}
\end{eqnarray*}
}}

for $y \in \{0,1\}.$ Here, $\Gamma(\pi(t)) = \pi(t)\mu _0 +
(1-\pi(t))\mu _1.$ If the agent uses the source in slot $t,$ and it
observes that the information is relevant, i.e., $A(t) = 1,$ and
$Z^y(t) = 1$ for any $y \in \{0,1\},$ then the state is known exactly
and $X(t) =1,$ thus belief $\pi(t+1) = \mu_1.$ Whereas if agent uses
the source, $A(t)=1$ but $Z^y(t) = 0$ then the state is known exactly
and $X(t) = 0,$ thus belief $\pi(t+1) = \mu_0.$ If the source is not
used, state is not observed but belief is updated.

From \cite{Bertsekas95a}, we know that the $\pi(t)$ captures the
information about the history $H_t$, and it is a sufficient
statistic. It suggests that the optimal policies can be restricted to
stationary Markov policies.  In this, one can obtain the optimum value
function by solving dynamic program.  We first define the
value function under initial action $A_1$ and availability $Y_1.$

{{
\begin{align*}
V_T := {} & \mbox{value function under $A_1=1, Y_1=1,$} \\
\widetilde{V}_T := {} & \mbox{value function under $A_1=1, Y_1=0,$} \\ 
V_{NT} := {} & \mbox{value function under $A_1=0, Y_1=1,$} \\
\widetilde{V}_{NT} := {} & \mbox{value function under $A_1=0, Y_1=0.$}
\end{align*} }}
We can write the following.
{{
\begin{dmath*}
V_T(\pi) =  \rho(\pi) + \beta [ (1-\pi )
 \{ {\theta ^1} V(\mu_1) + 
(1 - {\theta ^1}) \widetilde{V}(\mu_1)\} 
  + \pi\{ {\theta ^1} V(\mu_0) + (1 - {\theta ^1})
\widetilde{V}(\mu_0)\}] \\
{V_{NT}(\pi) = w + \beta [{\theta ^0}
V(\Gamma_1(\pi) ) + (1 - {\theta ^0}) }
\widetilde{V}(\Gamma_1(\pi))] 
\end{dmath*}
}}
{{
\begin{dmath*}
\widetilde{V}_T(\pi) =  \xi(\pi) + 
\beta [(1-\pi )\{ {\theta ^1}V(\mu_1) 
+ (1 - {\theta ^1})
\widetilde{V}(\mu_1)\} 
  + \pi\{ {\theta ^1}V(\mu_0) 
+ (1 - {\theta ^1})
\widetilde{V}(\mu_0)\}] \\
{\widetilde{V}_{NT}(\pi) =  w + \beta [{\theta ^0}V(\Gamma_0(\pi))
+ (1-{\theta ^0})\widetilde{V}(\Gamma_0(\pi))] }
\end{dmath*}
}}

Here $r(\pi) = (1 - \pi )r_1, \xi(\pi) = (1 - \pi )\eta_1, .$ The
optimal value function $V(\pi,y)$ and $\widetilde{V}(\pi,y),$ is
determined by solving the following dynamic program
{{
\begin{eqnarray*}
V(\pi) = \max \{ {V_T}(\pi),{V_{NT}}(\pi)\}; \texttt{ } 
 \widetilde{V}(\pi) = \max \{ {{\widetilde V}_T}(\pi),{{\widetilde
    V}_{NT}}(\pi)\}.
\end{eqnarray*} }}

\subsection{Structural Results}
We now derive structural results for value functions, convexity of
value functions and a threshold type policy. We will derive all result
for $\mu_0 > \mu_1.$ This means that source is positively correlated;
 a source that provides relevant information will more likely provide relevant information in future also.

\begin{lemma} \hspace{2cm}
\begin{enumerate}
\item For fixed $w$, $V(\pi),V_T(\pi),V_{NT}(\pi), \widetilde{V}(\pi),
  \widetilde{V}_{T}(\pi )$ and $\widetilde{V}_{NT}(\pi)$ are convex
  functions of $\pi.$
\item For a fixed $\pi$, $V(\pi),V_T(\pi),V_{NT}(\pi),
  \widetilde{V}(\pi), \widetilde{V}_{T}(\pi)$ and
  $\widetilde{V}_{NT}(\pi)$ are non decreasing and convex in $w.$
\item For fixed subsidy $w, \beta,$ and $\mu_0 > \mu_1,$ the value
  functions $V(\pi), V_{T}(\pi)$ and $V_{NT}(\pi)$ are decreasing in
  $\pi.$ Also, $\widetilde{V}(\pi), \widetilde{V}_{T}(\pi)$ and
  $\widetilde{V}_{NT}(\pi)$ are decreasing in $\pi.$
\item $(V_T(\pi) - V_{NT}(\pi))$ and $(\widetilde{V}_T(\pi) -
  \widetilde{V}_{NT}(\pi))$ are decreasing in $\pi.$
\end{enumerate}
\label{lemma:convex-pi-w} 
\end{lemma}
\begin{IEEEproof}
The proof of $(1)$ and $(2)$ is similar to the proof of Lemma 2 in \cite{Meshram16}.

3) The proof is done by induction technique. 
Assume that $V_n(\pi)$ and $\widetilde{V}_n(\pi)$ are non-increasing in $\pi.$ 
Let $\pi' \geq \pi$ and playing the arm is optimal.
Then,
\begin{dmath*}
V_{n+1}(\pi) = \rho(\pi) + \beta [ (1-\pi )
 \{ {\theta ^1} V_n(\mu_1) + 
(1 - {\theta ^1}) \widetilde{V}_n(\mu_1)\} \nonumber \\ 
  + \pi\{ {\theta ^1} V_n(\mu_0) + (1 - {\theta ^1})
\widetilde{V}_n(\mu_0)\}] 
\end{dmath*}
Here, $\rho(\pi)$ is decreasing in $\pi,$ i.e. $\rho(\pi')<\rho(\pi)$
for $\pi'>\pi.$ Hence,
\begin{dmath*}
V_{n+1}(\pi) \geq \rho(\pi') + \beta [ (1-\pi )
 \{ {\theta ^1} V_n(\mu_1) + 
(1 - {\theta ^1}) \widetilde{V}_n(\mu_1)\} \nonumber \\ 
  + \pi\{ {\theta ^1} V_n(\mu_0) + (1 - {\theta ^1})
\widetilde{V}_n(\mu_0)\}].
\end{dmath*}
From our assumption $\mu_0>\mu_1$, we get stochastic ordering on
observation probability, i.e., $[1-\pi, \pi]^T \leq_s [1-\pi',
  \pi']^T.$ and $V_n(\pi),\widetilde{V}(\pi)$ are decreasing in $\pi.$
Then we have
\begin{dmath*}
V_{n+1}(\pi) \geq \rho(\pi') + \beta [ (1-\pi' )
 \{ {\theta ^1} V_n(\mu_1) + 
(1 - {\theta ^1}) \widetilde{V}_n(\mu_1)\} \nonumber \\ 
  + \pi'\{ {\theta ^1} V_n(\mu_0) + (1 - {\theta ^1})
\widetilde{V}_n(\mu_0)\}]
\end{dmath*}
\begin{dmath*}
V_{n+1}(\pi) \geq V_{n+1}(\pi').
\end{dmath*}
Similarly we can show that $\widetilde{V}_{n+1}(\pi) \geq
\widetilde{V}_{n+1}(\pi').$ This is true for every $n \geq 1.$ From
Chapter~$7$ of \cite{Bertsekas95a} and Proposition~$2.1$ of
Chapter~$2$ of \cite{Bertsekas95a}, $V_n(\pi) \rightarrow V(\pi),$
uniformly and similarly $\widetilde{V}_n(\pi) \rightarrow
\widetilde{V}(\pi).$ Hence $V(\pi) \geq V(\pi')$ and
$\widetilde{V}(\pi) \geq \widetilde{V}(\pi')$ for $\pi' \geq \pi.$

Next we prove, $V_T(\pi)$ and $V_{NT}(\pi)$ is non increasing in $\pi.$
\begin{eqnarray}
V_T(\pi) =  \rho(\pi) + \beta [ (1-\pi )
 \{ {\theta ^1} V(\mu_1) + 
(1 - {\theta ^1}) \widetilde{V}(\mu_1)\} \nonumber \\ 
  + \pi\{ {\theta ^1} V(\mu_0) + (1 - {\theta ^1})
\widetilde{V}(\mu_0)\}] \\
V_{NT}(\pi) = w + \beta [{\theta ^0}
V(\Gamma_1(\pi) ) + (1 - {\theta ^0})
\widetilde{V}(\Gamma_1(\pi))] 
\end{eqnarray}
For $\pi_1 > \pi_2,$ 
\begin{dmath*}
V_T(\pi_1)-V_T(\pi_2) = (\pi_1-\pi_2)\beta \theta^1
(V(\mu_0)-V(\mu_1)) + (\pi_1-\pi_2) \beta
(1-\theta^1)(\widetilde{V}(\mu_0)-\widetilde{V}(\mu_1))
\end{dmath*}
Using above result, $V_T(\pi)$ is non increasing in $\pi.$
Similarly, $V_{NT}(\pi)$ is non increasing in $\pi.$ 

4) Let $D(\pi) = V_T(\pi) -V_{NT}(\pi)$ and $D(\pi)$ is decreasing in $\pi,$ i.e $D(\pi)<D(\pi')$ 
for $\pi>\pi'.$  We need to show 
\begin{dmath}
V_{T}(\pi)-V_{NT}(\pi) < V_{T}(\pi')-V_{NT}(\pi').
\label{eqn:submodular-I}
\end{dmath}
Rearranging~\ref{eqn:submodular-I}, 
\begin{dmath}
V_{T}(\pi)-V_{T}(\pi') < V_{NT}(\pi)-V_{NT}(\pi'). 
\label{eqn:submodular-II}
\end{dmath}
Now, the right hand side of the~\eqref{eqn:submodular-II},

\begin{dmath*}
V_{NT}(\pi)-V_{NT}(\pi') = \beta \theta^0
\{V(\Gamma_1(\pi))-V(\Gamma_1(\pi'))\} + \beta (1-\theta^0)
\{\widetilde{V}(\Gamma_1(\pi))-\widetilde{V}(\Gamma_1(\pi'))\} \\ \geq
\beta \theta^0 \{V(\mu_0)-V(\Gamma_1(\pi'))\} + \beta (1-\theta^0)
\{\widetilde{V}(\mu_0)-\widetilde{V}(\Gamma_1(\pi'))\} \\ \geq \beta
\theta^0 \{V(\mu_0)-V(\mu_1)\} + \beta (1-\theta^0)
\{\widetilde{V}(\mu_0)-\widetilde{V}(\mu_1)\}
\end{dmath*} 

The left hand side of the~\eqref{eqn:submodular-II},
\begin{dmath*}
V_{T}(\pi)-V_{T}(\pi') = (\rho(\pi)-\rho(\pi')) + \beta
(\pi-\pi')\theta^1 \{V(\mu_0)-V(\mu_1)\} + \beta
(\pi-\pi')(1-\theta^1) \{\widetilde{V}(\mu_0)-\widetilde{V}(\mu_1)\}
\end{dmath*}

Note that $\rho(\pi) - \rho(\pi') = r_1(\pi' - \pi) < 0$ because $\pi
> \pi'.$ Also, from the above expressions of difference in value
functions, we can easily see that that for $\theta^0 = \theta^1,$
Eqn~\eqref{eqn:submodular-II} is true.

Even for $\theta^0 \neq \theta^1,$ Eqn~\eqref{eqn:submodular-II} is
true because $\rho(\pi) - \rho(\pi') < 0$ and $0<\pi - \pi'<1.$ Hence,
$V_{T}(\pi)-V_{T}(\pi')<V_{NT}(\pi)-V_{NT}(\pi').$
Similar steps follow for
$(\widetilde{V}_T(\pi)-\widetilde{V}_{NT}(\pi)).$\\
\end{IEEEproof}

\begin{remark}
The above proofs have been done assuming that the availability
probability is independent of state and action. However, a similar
argument can be made for the dependent case $\theta^a(\pi,y)$ by
imposing following conditions. $\theta^a(\pi,1)>\theta^a(\pi,0),$ and
$\theta^a(\pi,y)>\theta^a(\pi ',y),$ for $\pi' > \pi.$
\end{remark}

We now define the threshold type policy and later we prove that the
optimal policy is threshold type.

\begin{definition}(Threshold type policy) 
A policy is said to be threshold type  if one of the following is true.
\begin{enumerate}
\item The optimal action is to play the arm for all $\pi.$
\item The optimal action is to not play the arm for all $\pi.$
\item There exists a threshold $\pi^*$ such that for all $\pi \leq \pi^*$
 the optimal action is to play the arm and not to play otherwise.
\end{enumerate}
\end{definition}
\begin{theorem}
For fixed $w$ and $\beta,$ 
\begin{enumerate}
\item The optimal policy is threshold type when arm is available,
  i.e., $\exists \pi^* \in [0,1]$ such that, $V_{T}(\pi)\geq
  V_{NT}(\pi)$ for $\pi \leq \pi_{th}$ and $V_T(\pi)<V_{NT}(\pi)$ for
  $\pi > \pi_{th}.$
\item The optimal policy is threshold type when arm is unavailable,
  i.e., $\exists \widetilde{\pi} \in [0,1]$ such that,
  $\widetilde{V}_{T}(\pi)\geq \widetilde{V}_{NT}(\pi)$ for $\pi \leq
  \widetilde{\pi}_{th}$ and $\widetilde{V}_T(\pi)<\widetilde{V}_{NT}(\pi)$
  for $\pi > \widetilde{\pi}_{th}.$
\end{enumerate}
\label{thm:threshold}
\end{theorem}
\begin{IEEEproof}
From Lemma~\ref{lemma:convex-pi-w}, the value functions
$V(\pi),\widetilde{V}(\pi)$ are convex in $\pi.$ Further, from
Lemma~\ref{lemma:convex-pi-w}.$5$ and
Lemma~\ref{lemma:convex-pi-w}.$6$ we know that
$V_T(\pi)-V_{NT}(\pi)$ and
$\widetilde{V}_T(\pi)-\widetilde{V}_{NT}(\pi)$ is decreasing with
$\pi$. This implies that there exists $\pi_{th},\widetilde{\pi}_{th} \in
[0,1]$ such that following is true
  \begin{enumerate}
 \item Either $V(\pi ) = V_T(\pi)$ for all $\pi \in
   [0,1]$ or $V(\pi ) = V_{NT}(\pi)$ for all $\pi \in [0,1]$ or 
 {{  
 \begin{eqnarray*} 
 V(\pi) = \begin{cases}
 V_T(\pi) & \mbox {  for $\pi \leq  \pi_{th},$ }  \\
 V_{NT}(\pi) & \mbox {  for $\pi \geq  \pi_{th}.$ }
\end{cases}
 \end{eqnarray*}}}
  \item Either $\widetilde{V}(\pi ) = \widetilde{V}_T(\pi)$ for all
    $\pi \in [0,1]$ or $\widetilde{V}(\pi ) = \widetilde{V}_{NT}(\pi)$
    for all $\pi \in [0,1]$ or there exists $\tilde{\pi}_{th}$ such
    that {{
 \begin{eqnarray*} 
 \widetilde{V}(\pi) = \begin{cases}
 \widetilde{V}_T(\pi) & \mbox {  for $\pi \leq  \widetilde{\pi}_{th},$ }  \\
 \widetilde{V}_{NT}(\pi) & \mbox {  for $\pi \geq  \widetilde{\pi}_{th}.$ }
 \end{cases}
 \end{eqnarray*}}}
 \end{enumerate}
\label{theorem:threshold-policy} 
Thus, the claims follows 
\end{IEEEproof}
\subsection{Indexability and Whittle index computation}
Recall that our interest is to seek the index type policy. We use
the threshold policy result to show indexability and later provide
an index computation algorithm. 

We now define indexability and index, it is motivated from
\cite{Whittle88,Dayanik02}.  Let $\mathcal{G}(w)$ be the subset of
state vector $S$ in which it is optimal to not play the arm with
subsidy $w,$ it is given as follows.
%
\begin{align}
\mathcal{G}(w) := & \{(\pi,y) \in [0,1] \times \{0,1\} : 
\nonumber \\
&
 V_T(\pi,w) \leq V_{NT}(\pi,w) ,
 \widetilde{V}_T(\pi,w) \leq \widetilde{V}_{NT}(\pi,w)\}.
\label{eq:G_w_set}
\end{align}
%
For clarity, we have explicitly mentioned dependence of value function
on $w.$ Using set $\mathcal{G}(w),$ indexability and index are
defined as follows.
\begin{definition}
An arm is indexable if $\mathcal{G}(w)$ is increasing in subsidy $w,$ i.e.,  
\begin{displaymath}
w_2 \le w_1 \Rightarrow \mathcal{G}(w_2) \subseteq \mathcal{G}(w_1).
\end{displaymath} 
\label{def:indexable}
\end{definition}

\begin{definition}
The index of an indexable arm is defined as 
\begin{equation}
w(\pi,y) := \inf \{w \in \mathbb{R}:(\pi,y) \in \mathcal{G}(w) ,
\forall (\pi,y) \in S\}.
\label{eqn:Whittle-index}
\end{equation}
\label{def:Whittle-index-a} 
\end{definition}

\begin{remark}\hspace{2cm}
\begin{enumerate}
\item Note that we can rewrite the definition of set $\mathcal{G}(w)$ in
  \ following way.
\begin{displaymath}
\mathcal{G}(w) = \left\{ [\pi_{th},1] \times \{1\}, [\widetilde{\pi}_{th},1]
\times \{0\} \right\},
\end{displaymath}
where $\pi_{th}:= \min \{ \pi \in [0,1] :V_T(\pi,w) \leq V_{NT}(\pi,w)\},$ and
$\widetilde{\pi}_{th}:= \min \{ \pi \in [0,1] :\widetilde{V}_T(\pi,w) \leq
\widetilde{V}_{NT}(\pi,w)\}.$
If the optimal policy is of threshold type, then $\pi_{th}$ and
  $\widetilde{\pi}_{th}$ are singleton.
\item Here, the definition of indexability and index is motivated from
  work of \cite{Whittle88} on restless bandits. In standard restless
  bandits, arms are assumed to be always available and $y = 0$ is not
  feasible option. 
\item When $\theta^{a} = 0$ or $\theta^{a} = 1$ for all $a \in
  \{0,1\},$ our definitions of indexability and index are still valid.
\end{enumerate}
\end{remark}

To claim indexability, we will require to show that $\pi_{th}(w)$ and
$\widetilde{\pi}_{th}(w)$ are non-increasing in $w.$
Now, we use the following lemma from~\cite{Meshram16}.

\begin{lemma} 
 If
{\small{
\begin{eqnarray*}
\frac{\partial V_T(\pi,w)}{\partial w}\bigg|_{\pi=\pi_{th}(w)} &<&
\frac{\partial V_{NT}(\pi,w)}{\partial w}\bigg|_{\pi=\pi_{th}(w)},
\\ 
\frac{\partial \widetilde{V}_T(\pi,w)}{\partial
  w}\bigg|_{\pi=\widetilde{\pi}_{th}(w)} &<& \frac{\partial
  \widetilde{V}_{NT}(\pi,w)}{\partial
  w}\bigg|_{\pi=\widetilde{\pi}_{th}(w)},
\end{eqnarray*} }}
then $\pi_{th}(w)$ and $\widetilde{\pi}_{th}(w)$ are monotonically decreasing
functions of $w.$
\label{lemma:indexibility}
\end{lemma} 
Now, using Lemma~\ref{lemma:indexibility} and
Definition~\ref{def:indexable}, we can show that single-armed restless
bandit is indexable.
\begin{theorem}
If $\mu_0>\mu_1$ and $\beta < 1/3,$ then a single-armed restless
bandit is indexable.
\end{theorem}
\begin{IEEEproof}
 The following inequalities obtain using induction technique, .
{{\begin{equation*}
\bigg\vert \frac{\partial V(\pi, w)}{\partial w } \bigg \vert, 
\bigg\vert \frac{\partial V_T(\pi, w)}{\partial w } \bigg \vert,
\bigg\vert \frac{\partial V_{NT}(\pi, w)}{\partial w } \bigg \vert
\leq \frac{1}{1-\beta}
\end{equation*}
}}
and
{{
\begin{equation*}
\bigg\vert \frac{\partial \widetilde{V}(\pi, w)}{\partial w } \bigg \vert, 
\bigg\vert \frac{\partial \widetilde{V}(\pi, w)}{\partial w } \bigg \vert,
\bigg\vert \frac{\partial \widetilde{V}_{NT}(\pi, w)}{\partial w } \bigg \vert
\leq \frac{1}{1-\beta}
\end{equation*}
}}
Also,
{{
\begin{dmath*}
\frac{\partial V_T(\pi, w)}{\partial w } = \beta \left[
  (1-\pi) \{ \theta^1 \frac{\partial V(\mu_1, w)}{\partial w } +
  (1-\theta^1) \frac{\partial \widetilde{V}(\mu_1, w)}{\partial w
  }\} + \pi \{ \theta^1 \frac{\partial V(\mu_0, w)}{\partial w } +
  (1-\theta^1) \frac{\partial \widetilde{V}(\mu_0, w)}{\partial w
  }\} \right]
\end{dmath*}
}}
and
{{
\begin{dmath*}
\frac{\partial V_{NT}(\pi, w)}{\partial w } = 1 + \beta \{ \theta^0 \frac{\partial V(\Gamma_1(\pi),w
  )}{ \partial w } + (1-\theta^0) \frac{\partial V(\Gamma_1(\pi),w
  )}{ \partial w }\}.
\end{dmath*}
}} 
Now from Lemma~\ref{lemma:indexibility}, we require the difference
$\frac{\partial V_{NT}(\pi, w)}{\partial w } - \frac{\partial V_T(\pi,
  w)}{\partial w } $ to be non-negative at $\pi_{th}(w)$ and
$\widetilde{\pi}_{th}(w)$ . This reduces to following expression.
{{ 
\begin{dmath}
 \left[
  (1-\pi) \{ \theta^1 \frac{\partial V(\mu_1, w)}{\partial w } +
  (1-\theta^1) \frac{\partial \widetilde{V}(\mu_1, w)}{\partial w
  }\} + \pi \{ \theta^1 \frac{\partial V(\mu_0, w)}{\partial w } +
  (1-\theta^1) \frac{\partial \widetilde{V}(\mu_0, w)}{\partial w
  }\} \right] - {\left[ \theta^0 \frac{\partial V(\Gamma_1(\pi),w
  )}{ \partial w } + (1-\theta^0) \frac{\partial V(\Gamma_1(\pi),w
  )}{ \partial w } \right] <   \frac{1}{ \beta}.}
\label{eqn:nec-cond}
\end{dmath}
}}
We can provide upper bound on LHS of above expression and it is upper
bounded by $2/(1-\beta).$ If $\beta < 1/3,$ Eqn.~\eqref{eqn:nec-cond}
is satisfied. $\pi_{th}(w)$ is decreasing in $w.$ Similarly
$\widetilde{\pi}_{th}(w)$ is decreasing in $w.$ And the claim follows.
\end{IEEEproof}
Proof of indexability for $0<\theta^{a} <1$ requires assumption on
$\beta$. Whereas for $\theta^a = 1$ or $0,$ indexability do not need
assumption on $\beta,$ because the value function expression can be
easily derived and then differentiating w.r.t. subsidy $w$, we can get
required, such result is studied in \cite[Theorem~1]{LiuZhao10}.
%
We now use Definition~\ref{def:Whittle-index-a} and restate the
Whittle index definition as follows.

\begin{definition}[Whittle's index]
For a given belief $\pi \in [0,1]$ and availability $y \in \{0,1\},$
Whittle index $w(\pi,y)$ is the minimum subsidy $w$ for which not
playing the arm is the optimal action.
\begin{eqnarray}
w(\pi,1) = \inf \{ w \in \mathbb{R}: V_{NT}(\pi) = V_T(\pi)\},
\nonumber \\ 
w(\pi,0) = \inf \{ w \in \mathbb{R}: \widetilde{V}_{NT}(\pi) = 
\widetilde{V}_T(\pi)\}.
\end{eqnarray}
\end{definition}

When $\theta^a= 0,1$ for all $a \in \{0,1\},$ the expression for index
can be computed and this is given in \cite[Section IV]{LiuZhao10}. But
for $\theta^a \in (0,1),$ it is very difficult to obtain closed form
expression for value functions because there is coupling between
action value functions.

Hence, we study a numerical scheme for Whittle index computation.  This
scheme uses the threshold result of value functions and two-timescales
stochastic approximations. In two-timescales stochastic
approximations, we update $w_t$ at slower timescales or natural
timescales, and the value functions are updated using value iteration
algorithm at faster timescales. This scheme here is inspired from
stochastic approximation algorithms, see \cite{Borkar08, Borkar17}.

In this scheme for fixed $w,$ $y=1$ and a threshold $\pi,$ we know
that $V_{T}(\pi,w)= V_{NT}(\pi,w).$ Using value iteration algorithm,
we compute $V_{T}(\pi,w)$ and $V_{NT,w}(\pi,w)$ on faster time scales
until difference $|V_{T}(\pi,w)- V_{NT,w}(\pi,w)|$ becomes smaller
than tolerance $h.$ To compute the index $w(\pi,1),$ our algorithm
starts with initial subsidy $w_0$ and it is updated iteratively at
slower timescales according to following expression.
\begin{eqnarray*}
w_{t+1} = w_t + \alpha(V_T(\pi,w_t)-V_{NT}(\pi,w_t)).
\end{eqnarray*}
These computations are performed till difference $|V_{T}(\pi,w_t)-
V_{NT}(\pi,w_t)|$ is smaller than tolerance $h.$

Using similar procedure mentioned above, we update $w_{t}$ with slower
timescales and run value iteration for $\widetilde{V}_T(\pi,w_t)$ and
$\widetilde{V}_{NT}(\pi,w_t)$ on faster timescales when $\pi$ is
threshold and $y=0$. Hence this is used to compute the index
$w(\pi,0).$ The details are given in Algorithm~\ref{algo:WI}.  The
convergence of two timescales stochastic approximation algorithm is
presented in \cite[Chapter $6$]{Borkar08}.
\begin{algorithm}
\DontPrintSemicolon 
\KwIn{Reward values $r_1,\eta_1$; Initial subsidy $w_0,$
  tolerance $h,$ step size $\alpha.$ } 
\KwOut{Whittle index,$w(\pi,y)$}
\eIf {(y==1)}{ $w_t \gets w_0$\; 
  \While
  {$|V_T(\pi,w_t)-V_{NT}(\pi,w_t)|>h$ } { $w_{t+1} = w_t +
    \alpha(V_T(\pi,w_t)-V_{NT}(\pi,w_t)); $\\ $t=t+1;$\\ compute
    $V_T(\pi,w_t)$, $V_{NT}(\pi,w_t);$  } }{ $w_t \gets w_0$\; 
  \While
  {$|\widetilde{V}_T(\pi,w_t)-\widetilde{V}_{NT}(\pi,w_t)|>h$ } { $w_{t+1}
    = w_t + \alpha(\widetilde{V}_T(\pi,w_t)-\widetilde{V}_{NT}(\pi,w_t));
    $\\ $t=t+1;$\\ compute $\widetilde{V}_T(\pi,w_t)$,
    $\widetilde{V}_{NT}(\pi,w_t);$  } } 
\Return{$w(\pi,y)=w_t$}\;
\caption{ Algorithm that computes Whittle index for the single
  arm}
\label{algo:WI}
\end{algorithm}
\section{Numerical Results}
\label{sec:numericals}
We now evaluate performance of index policy and myopic policy. The
algorithms included in the comparative analysis are 1) Whittle's index
policy (WI)---contacts $M$ sources with highest Whittle's indices, 2)
myopic policy (MP)---contacts sources according to their expected
immediate rewards, 3) uniform random policy(UR)---chooses randomly
with uniform distribution.

Simulations were performed using MATLAB. In these simulations, the
sources start in random states and random initial beliefs. The initial
availability of sources are random. The reward is accumulated at the
end of each slot from sources that are contacted. These rewards are
stored and averaged over large number of iterations.

We consider example of an agent with $N=15$ sources and use following
parameters.

{{
\begin{eqnarray*}
\mu_o = [0.9, 0.9, 0.9, 0.9, 0.9, 0.9, 0.9, 0.9, 0.9, 0.9, 
 0.66, 0.69, 0.75, 0.78, 0.87],\\
\mu_1 = [0.1, 0.1, 0.1, 0.1, 0.1, 0.1, 0.1, 0.1, 0.1, 0.1, 
0.28, 0.25, 0.2, 0.15, 0.1], \\
r_1 = [1.25, 0.8, 0.8, 0.8, 0.8, 1.5, 1.3, 1.25, 1.2, 1.2, 
 1.4, 1.3, 1.2, 1.35, 1.15],\\
\eta_1 = [0.7, 0.7, 0.7, 0.7, 0.75, 0.7, 0.7, 0.7, 0.7, 0.7,
 0.9, 0.8, 0.7, 0.6, 0.7],\\
\theta^0 = [1, 1, 1, 1, 1, 0.35, 0.45, 0.75, 0.85, 0.9, 0.35,  
 0.45, 0.75, 0.85, 0.9],\\
\theta^1 = [1, 1, 1, 1, 1, 0.35, 0.45, 0.75, 0.85, 0.9, 
 0.35, 0.45, 0.75, 0.85, 0.9].
\end{eqnarray*}
}}
At each decision making instant the agent chooses to contact $M$ of
$N$ sources, where  we use $M=3.$ Our parameter set
represents the scenario where availability of sources is independent
of agent's decision but the perception of their usefulness depends on
it. Further, the sources in our examples tend to maintain their
information quality (relevant or irrelevant). 
\begin{figure}
  \begin{center}
    \begin{tabular}{cccc}
      \includegraphics[scale=0.24]{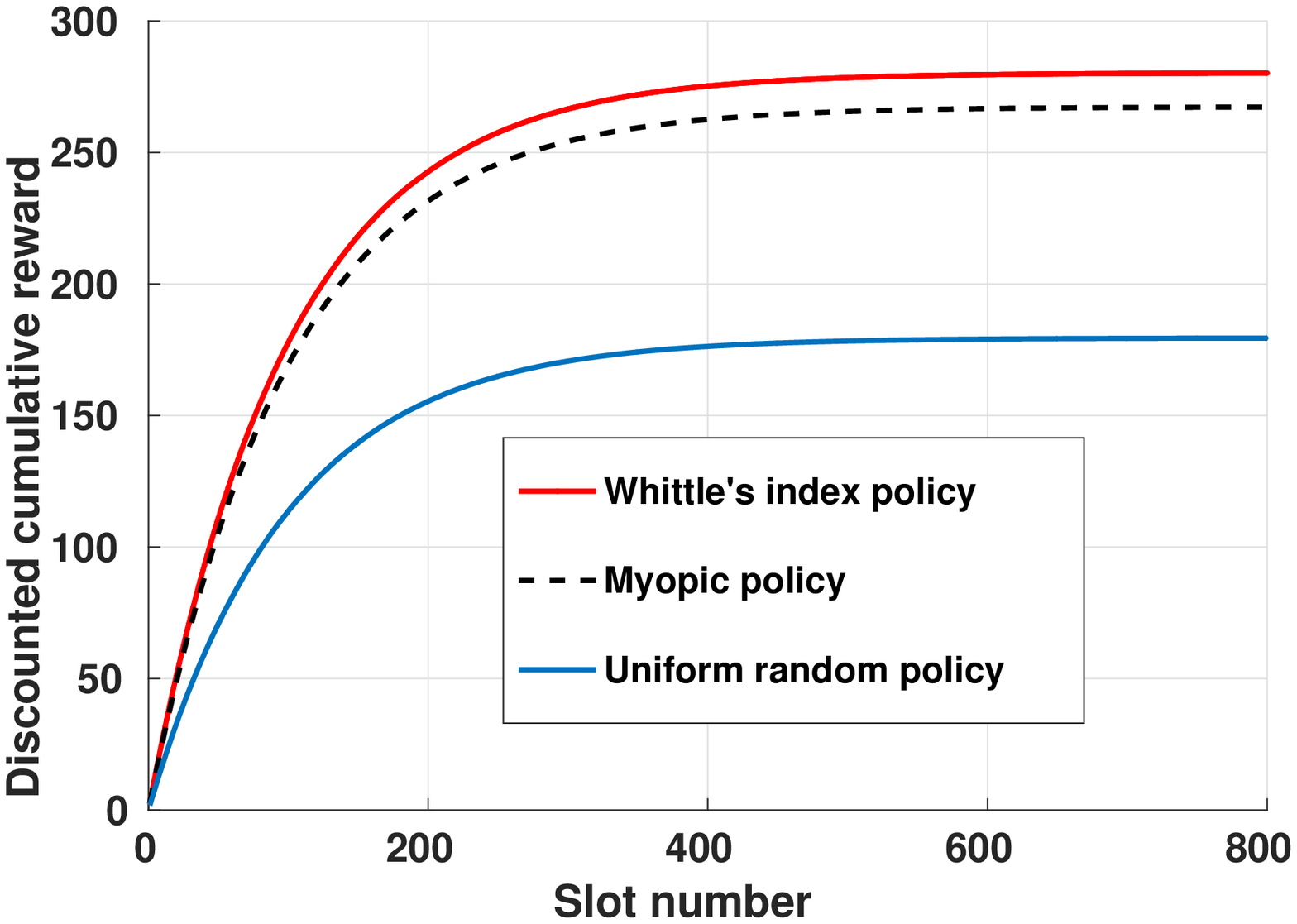}
      & 
      \hspace{-0.7cm}
      \includegraphics[scale=0.24]{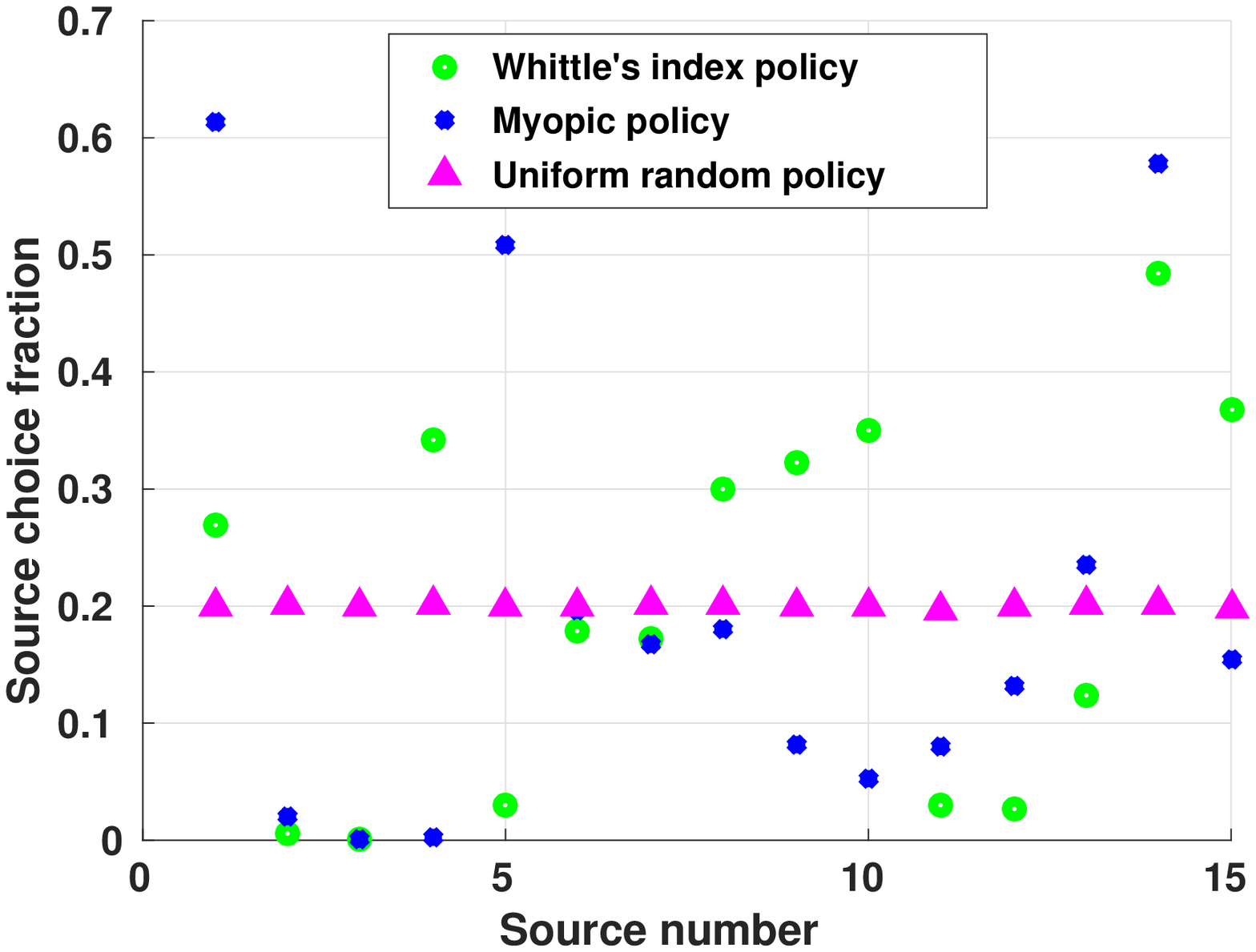} \\
      a) {\small{ Discounted cumulative reward}}   & \hspace{-0.7cm} b) {\small{ Source choice fraction }}  
     
    \end{tabular}
  \end{center}
  \caption{ We plot a) discounted cumulative rewards as function of
    time slot for different policies and b) source choice fraction for
    different policies. }
  \label{fig:reward-armchoice-addEx1}
\end{figure}
In Fig.~\ref{fig:reward-armchoice-addEx1}-a we plot the discounted
cumulative reward as a function of time slots. It can be seen that the
discounted cumulative reward under Whittle index policy (WI) is
comparable with that of the myopic policy (MP). We also observe that
performance of WI policy yields higher discounted cumulative reward
compared to that of myopic policy and uniform random policy. To gain
insight, we also plot source choice fraction as function of number of
arms in Fig.~\ref{fig:reward-armchoice-addEx1}-b. and which is the
probability that an source/arm is chosen in a slot.  Here, myopic
policy contacts sources $\{1,5,14\}$ most frequently compared to other
sources and this is because sources $\{1,5\}$ are always available and
$14$ has high reward. Whereas WI policy contacts from broader set of
sources more frequently even though they have lesser rewards.  This
behavior of Whittle's index policy is because it considers future
rewards and availability of sources through the action value
functions. Interestingly, sources $\{9,10,15\}$ that are not always
available are chosen.

\section{Concluding Remarks}
We formulated problem of information gathering in a social network
with dynamic availability of sources and time varying information
quality using RMAB model.  We studied the Whittle's index
policy. Also, the performance of this policy was illustrated for
moderate sized scenarios. 

In this work we considered the decision model of a single agent in a
social or information network.  
This can be used to model an 
individual element in a larger framework for studying information
acquisition and dissemination in social networks. For example, one may
consider the impact of a set of compromised or fake news sources over
the decisions of various agents across a network.

\bibliographystyle{IEEEbib} 
\bibliography{varun}

\end{document}